# MaskAnyone Toolkit: Offering Strategies for Minimizing Privacy Risks and Maximizing Utility in Audio-Visual Data Archiving

*Completed Research Paper (Under Review)*

Babajide Alamu Owoyele[1,4], Martin Schilling[2], Rohan Sawahn[2], Niklas Kaemer[2], Pavel Zherebenkov[2], Bhuvanesh Verma[2], Wim Pouw[3], Gerard de Melo[2]

## Abstract

This paper introduces MaskAnyone, a novel toolkit designed to navigate some privacy and ethical concerns of sharing audio-visual data in research. MaskAnyone offers a scalable, user-friendly solution for de-identifying individuals in video and audio content through face-swapping and voice alteration, supporting multi-person masking and real-time bulk processing. By integrating this tool within research practices, we aim to enhance data reproducibility and utility in social science research. Our approach draws on Design Science Research, proposing that MaskAnyone can facilitate safer data sharing and potentially reduce the storage of fully identifiable data. We discuss the development and capabilities of MaskAnyone, explore its integration into ethical research practices, and consider the broader implications of audio-visual data masking, including issues of consent and the risk of misuse. The paper concludes with a preliminary evaluation framework for assessing the effectiveness and ethical integration of masking tools in such research settings.

## Introduction

Audio-visual data with human subjects is crucial for behavioral sciences and linguistics and provides insights into human behavior and communication(Abney et al. 2018; Cienki 2016; D'Errico et al. 2015; Gregori et al. 2023). However, including identifiable human data raises ethical and privacy concerns(Abay et al. 2019; Benson et al. 2020; Bishop 2009; Jarolimkova and Drobikova 2019; Jeng et al. 2016; Johannesson and Perjons 2021). GDPR outlines frameworks for handling such data responsibly(Nautsch et al. 2019). In the spirit of open science, more platforms, artifacts, and tools are needed to balance privacy and data sharing, especially in social sciences and humanities (Hunyadi et al., 2016; Qian et al., 2018). Using audio-visual data in social and behavioral sciences requires a deep understanding of ethical, legal, and methodological aspects. However, using such data is necessary to ensure research transparency and reproducibility, with recent work arguing that retaining interview data should be the default(Resnik et al. 2024). Incorporating audio-visual data in research introduces a host of complexities, from data collection and analysis to interpretation. This can impact the reliability and validity of findings, raising concerns about potential biases, misinterpretations, and inadvertent capture of unrelated audiovisual data. Aligning and harmonizing audio-visual inputs is particularly challenging, especially with recent trends in generating audio-visual data. Considering these challenges, the proposed toolkit, MaskAnyone, becomes a necessity. Researchers must take the lead in considering the unique challenges associated with audio-visual data to safeguard their research's integrity, validity, and ethical conduct.

---

[1] Artificial Intelligence and Intelligent Systems, Hasso Plattner Institute, Potsdam, Berlin-Brandenburg, Germany, babajide.owoyele@hpi.de

[2] Hasso Plattner Institute, University of Potsdam, Potsdam, Berlin-Brandenburg, Germany.

[3] Donders Centre for Cognition, Radboud University, Nijmegen, Netherlands, wim.pouw@donders.ru.nl

[4] Dutch Research Institute for Transitions, Erasmus University Rotterdam, Netherlands, owoyele@drift.eur.nl

Building on existing work (Khasbage et al. 2022; Owoyele et al. 2022), we propose MaskAnyone, a toolkit for de-identifying individuals using audio-visual data. MaskAnyone offers multi-person masking, real-time bulk processing, and a user-friendly interface[3] - essential for handling large datasets efficiently while reducing privacy risks. With accessibility in mind, social scientists can use MaskAnyone techniques to mask identifiable information, such as face-swapping and auditory elements. This modular approach allows researchers who do not write code to customize the anonymity level based on the sensitivity of the data and available computing resources. The toolkit's flexibility is essential as it enables researchers to use it on personal machines for smaller projects or scale it up to server-based environments for institutional research involving larger datasets. This adaptability allows researchers to tailor the toolkit to their needs and resources. Considering the challenges and related issues above, as well as the opportunities to leverage developments in information systems and computer science, our paper is guided by the following research questions and objectives:

1. How can researchers effectively navigate/balance subjects' privacy with the utility of audio-visual data?
2. What techniques can be employed to ensure the ethical use of audio-visual data in compliance with stringent regulatory frameworks?
3. What implications does masking audio/visual data have for generating and analyzing more synthetic data, and what evaluation-related challenges remain in designing and iterating such tools?

Drawing explicitly on the design science research framework(Hevner et al. 2004), we iteratively developed *a scalable toolkit to de-identify audiovisual data. Co-developed with researchers and data stewards, the tool can be integrated into current research practices to promote ethical data sharing*. MaskAnyone aims to navigate the privacy risks associated with audio-visual data sharing. Distinct from existing solutions (Khasbage et al. 2022; Owoyele et al. 2022), the toolkit also supports exporting body and face pose data as JSON and Csvs along with *multi-person masking* and *real-time bulk processing* while offering *a modular, user-friendly interface*. It incorporates various masking techniques for visual (e.g., face-swapping) and auditory elements of identifiable information and is designed to be scalable for handling large datasets. By providing customizable options, MaskAnyone allows researchers to balance privacy and data utility, and it is specifically designed for scalability. Tailoring to data sensitivity requirements and computing needs, MaskAnyone can run on personal machines or a (secured) server with a large amount of computing power to accommodate multiple researchers with large datasets. Our paper proceeds as follows: First, we delve into recent literature on the masking of audiovisual data and discuss the design methodology used. Next, we detail the techniques and algorithms we employed for de-identification in video and audio domains. We present our preliminary evaluations, describing the system, sample datasets, and proposed evaluation metrics. The discussion section offers a broader outlook on how we want to evaluate toolkits like MaskAnyone, ending with broader implications, opportunities, and issues of masking practices.

## Related Work/Problem Space

Masking has become even more relevant due to the development of MaskAnyone, with several implications that need to be considered. One of the issues that need to be addressed is consent. Regulatory frameworks like GDPR and HIPAA require that human subjects be informed about de-identification techniques. Therefore, masking previously collected data may require additional consent. MaskAnyone is designed to safeguard privacy, but there is also potential for misuse by enabling the creation of deepfakes. However, it might also generate data to detect deepfakes. There are also questions about what degree of masking is sufficient and how to determine it. For instance, sensitive contexts like healthcare may require complete

---

[3]https://anonymous.4open.science/r/Privacy4MultimodalAnalysis-57D4/results/masking_s2_masking.png

*MaskAnyone Toolkit*anonymization and additional safeguards to enable audio-visual data sharing. As a field, we need to be able to judge and validate what degree of masking is considered a reduction of identifiable information, de-identification, pseudo-anonymization, or complete irreversible anonymization. This question is not just an abstract theoretical question, as failure to comply with regulatory privacy frameworks can have direct legal repercussions. These considerations form the challenging context within which audio-visual masking tools must operate. Nonetheless, when integrated into ethical procedures already upheld by research institutes, toolkits like MaskAnyone promise to support the safe sharing of multimodal language data.

## *Person de-identification in Video*

For video de-identification (Gafni et al. 2019; Kasturi and Ekambaram 2014), we distinguish two principal strategies: hiding and masking. The hiding strategy focuses on obscuring or eliminating video segments containing personally identifiable information, thereby enhancing privacy. In contrast, the masking strategy substitutes the individual with an alternate representation, preserving essential attributes for the video's utility while masking the individual's identity. Recently, tools such as Masked-Piper (Owoyele et al. 2022) and Red Hen Anonymizer (Khasbage et al. 2022) have been developed to use hiding and masking strategies; however, their usability remains challenging, and they lack modularity, adding new modules for privacy masking and multimodal analytics. Regardless of the strategy employed, a preliminary step of person detection is needed. This can be achieved through generic object detection models like 'You Only Look Once' (YOLO) (Redmon et al. 2016) or more specialized models like MediaPipe's BlazePose (Bazarevsky et al. 2020). Both categories of models offer robust capabilities, catering to a range of requirements and applications.

**Hiding**

Upon detecting a person in a video, the hiding strategy offers multiple options for concealment. One common approach is to obfuscate the identified area using techniques such as Gaussian blurring, pixelation, or Laplacian edge detection, which can be implemented using image processing libraries such as OpenCV (Bradski 2000; OpenCV 2023). While machine learning-based methods such as PiDiNet(Su et al. 2021) and UAED (Zhou2023) may offer enhanced quality and privacy, they do not necessarily guarantee robust anonymity, particularly for videos featuring well-known individuals (Hasan et al. 2018; Lander et al. 2001). Another tactic involves overlaying the detected area with a non-translucent color, effectively hiding person-specific attributes except for potentially identifiable characteristics like height or clothing. Alternatively, one could employ inpainting techniques to estimate and replace the background behind the detected individuals, as demonstrated by projects such as Inpaint Anything (Yu et al. 2023), STTN(Zeng et al. 2020), and E2FGVI  (Dang and Buschek 2021; Ripperda et al. 2020)(Li2022). While earlier methods like STTN had limitations such as resolution constraints, recent developments like E2FGVI offer more flexibility and improved performance. The field continues to evolve, mainly focusing on responsible data sharing (Morehouse 2023), with emerging approaches like DMT (Yu 2023-2) promising even more effective results.

**Masking**

Hiding leads to a high loss of information, such as facial expressions or other expressive movements, that may be important for linguistic and behavioral research. Masking aims to maintain a representation of this information without retaining personally identifiable information. **Landmark Detection** for humans in monocular video data refers to identifying and localizing specific points on the human body, such as joints or facial features. The task can be further divided into 2D and 3D Landmark detection. Several models offer advanced features for landmark detection but have specific limitations -- be it computational speed, accuracy, or full-body image handling. For example, OpenPose is noted for its accuracy but is computationally slower(Mroz et al. 2021). AlphaPose, an open-source model, has gained attention for innovative approaches like Symmetric Integral Keypoints Regression (Fang2022). ViTPose offers a scalable architecture for human pose estimation based on a Vision Transformer that can scale up to 1B parameters (Xu et al. 2022).  MediaPipe's Holistic model was developed to mitigate limitations in full-body detection and offers an approximation of 3D positions. We, therefore, settled on using this approach for the current version of MaskAnyone. The BlazePose model underlying MediaPipe uses a two-stage approach, in which a single-shot-detection (SSD) based detector first locates the bounding boxes of people (Grishchenko and

*Forty-Fifth International Conference on Information Systems, Bangkok, Thailand 2024*
**3**

Bazarevsky 2020). Subsequently, an estimation model applies a regression approach supervised by a combined heat map/offset prediction of all key points.

MediaPipe's holistic model computes fine-grained landmarks for the complete body (including the pose, detailed hand landmarks, and face mesh). To do so, pose detection is performed first. Subsequently, a region-of-interest (ROI) detection is performed to locate the hands and face based on the detected pose. A unique transformer re-crop model is then used to improve the ROI at only 10% of the corresponding model's inference time (Grishchenko and Bazarevsky 2020). Finally, these crops are passed to the detailed hand landmarks or face mesh models. **Face swapping**, widely used in entertainment and social media, replaces one person's face with another in images or videos. Despite its potential misuse in creating malicious deep fakes, it can serve as a de-identification tool by swapping faces to obscure identities while retaining expressions. This approach is implemented in the Red Hen Anonymizer using an FSGAN-based method (Nirkin et al. 2019). DeepFakes and DeepFaceLab offer pipelines for face replacement but often require post-editing for natural results and are limited by their need for retraining for each new face pair (Korshunov and Marcel 2018; Liu et al. 2023). InsightFace could face-swap without retraining but turned commercial and withdrew public models (InsightFace 2023). The Roop project, based on InsightFace, was discontinued due to ethical concerns. Other methods, such as Thin-Plate Spline and XFace, are limited in preserving facial expressions or are unsuitable for large-scale de-identification (Balci 2005; Zhao and Zhang 2022). Facebook's de-identification approach uses a feed-forward network, maintains facial features, and does not require retraining, but it is not open-source (Gafni et al. 2019). **Avatar creation** focuses on modeling the human body in 3D, including texture generation. Open-source tools like Blender offer basic capabilities for avatar creation (cgtinker 2021). On the proprietary side, solutions such as Meshcapade, Rokoko, and UnrealEngine MetaHuman Animator provide high-quality results. Still, they are not open-source and can be costly for research (Meshcapade 2023; Rokoko 2023). Our experiments with Meshcapade revealed a discrepancy between the advertised and actual quality.

### *Person de-identification in Audio*

Voice data inherently contains personally identifiable information through spoken content, vocal attributes, or linguistic style (Tomashenko et al. 2020). Our work zeroes in on techniques for obscuring identifiable information while retaining linguistic and prosodic elements of speech. Broadly, we explore Spectral Modification, Pitch Shifting, and Voice Conversion as standard methodologies. Spectral Modification alters speech signals at the spectral level, targeting formant frequencies or the spectral envelope. Pitch Shifting directly modifies pitch but can distort speech naturalness and paralinguistic elements. On the other hand, Voice Conversion adapts one speaker's vocal characteristics to resemble another's, offering a nuanced approach to de-identification. The Voice Privacy Challenge (Tomashenko et al. 2020) aims to advance voice anonymization techniques. The 2020 iteration offered two baselines: one using x-vectors (Fang et al. 2019) and neural speech synthesis and another altering voice signal through a McAdams Coefficient method (Patino et al. 2020). The 2022 round (Tomashenko et al. 2022), introduced improved baselines and evaluation metrics, employing the equal error rate (EER) for privacy and word error rate for utility. A noteworthy trade-off between privacy and utility was observed. For instance, submission T04 achieved a high EER of 47.60% but had a low pitch correlation of 37%, limiting its utility. Conversely, submission T18 balanced this trade-off well, boasting an 82% pitch correlation and an EER of 20.8%. Voice changer systems can be organized into four primary taxonomies: Phonetic Models employ phonemes for voice conversion, typically initiating the process by extracting phonemes from the input speech. Statistical Models, such as Gaussian Mixture Models (Reynolds 2015) or Hidden Markov Models (Kong et al. 2020; Wang et al. 2020), can be used to represent vocal features statistically and then map them to a target speaker. Deep Learning Models utilize architectures like CNNs, RNNs, and GANs, including notable examples such as HifiGAN and StarGAN, which generate high-quality converted speech using GAN architectures (Kong2020, Wang2020). Lastly, Retrieval-based Models search and concatenate similar speech segments from a target speaker's database. HifiGAN and StarGAN use GAN-based methods to synthesize speech from mel-spectrogram features. StarGAN also offers the potential for one-shot voice conversion (Pavlakos et al. 2019). VITS incorporates a Conditional Variational Autoencoder coupled with adversarial learning for end-to-end speech synthesis (Kim et al. 2021). The Retrieval-based Voice Conversion system employs feature extractors like Crepe for F0 features and HuBERT for representing the input speech (Hsu et al. 2021; Kim et al. 2018). These are then synthesized into the final vocal output through models such as HiFiGan (Kong et al. 2020). Notably, the literature still lacks explicit benchmarks that evaluate these models regarding utility preservation and privacy assurance. This is understandable as these individual tools were not explicitly developed for our context and problem space – to help social and behavioral science researchers navigate audio-visual data-sharing practices and the risk inherent in such endeavors.

## Methodology - Design Science Research

We adopted the design science methodology for designing MaskAnyone. Here, we elaborate on the design requirements guiding the toolkit construction by integrating the design science approach and insights from the literature on audio-visual data sharing in social and behavioral sciences. This ensures that MaskAnyone augments ethical, robust, and effective data management and sharing practices. We have also developed MaskAnyone to balance open science and privacy in audio-visual research based on the call for more FAIR-friendly research infrastructure and tools to support the development of thematic digital competence centers("Roadmaps from the Three Thematic DCCs – Digital Competence Centres | NWO" n.d.; "Thematic Digital Competence Centers | NWO" 2024). In terms of an artifact and as a DSR instantiation, the toolkit is modular and extendable and enhances the capabilities of existing tools by supporting advanced features like 3D tracking and real-time processing. It offers various masking methods for sharing and secure storage scenarios, thus protecting against unauthorized access risks like data breaches. MaskAnyone meets diverse research needs while ensuring data integrity and privacy by providing an easy-to-use interface and versatile masking options.[4]

### *Design Requirements*

Our toolkit, MaskAnyone, is developed using the Design Science Research framework as an instantiation. (Hevner et al. 2004). We enumerate the following design requirements based on user feedback from two live demo workshops with researchers at a Dutch University(n=16), at a German University (n=10), and data stewards(n=5). We have also used insights from existing literature on multimodal analysis to justify the relevance of these requirements, although we know they are non-exhaustive depending on future scenarios and evolving stakeholder needs (in multimodal behavior research)

| Requirement | Description | Source/Literature |
|---|---|---|
| R1 - Multimodality | Enable the masking of primary actors, backgrounds, and voice data to ensure comprehensive privacy. | Workshop/Survey |
| R2- Usability | Toolkit usability: intuitive enough for researchers with limited technical skills for quick deployment. | Workshop/Survey |
| R3- Flexibility | Flexibility in masking methods and settings to handle various video types and privacy constraints. | Workshop |
| R4- Multi-level | Offer different levels of de-identification to match specific use-cases and privacy needs. | Workshop/Survey |
| R5- Efficiency | Ensure performance efficiency on local machines, respecting hardware limitations. | Survey |
| R6- Sensitivity | Optimize performance for server environments, considering data sensitivities. | Survey |

---

[4]https://anonymous.4open.science/r/Privacy4MultimodalAnalysis-57D4/results/masking_s2_masking.png

| | | |
|---|---|---|
| R7- Scalability | Scalability to support distributed video processing for large datasets. | Workshop |
| R8- Modularity | Extensibility to integrate new techniques seamlessly as they emerge in video processing. | Workshop/Survey |

**Table 1. MaskAnyone Design Science Approach and Requirements**

# Solution-Space: MaskAnyone Architecture and User Interaction Layout

## *Architecture*

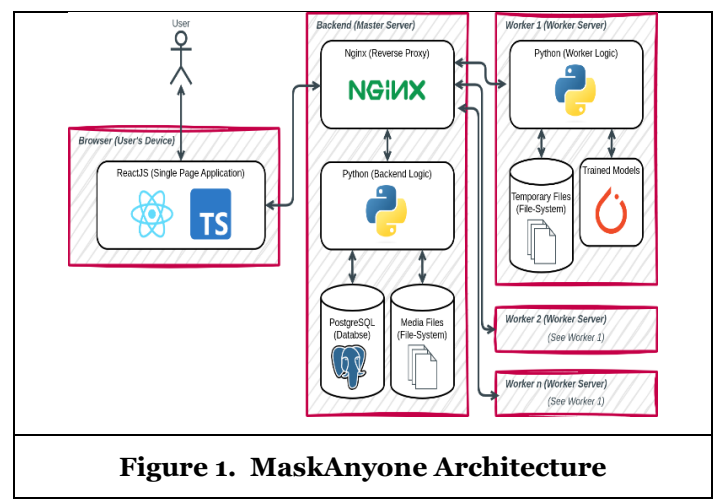

**Figure 1. MaskAnyone Architecture**

In response to the requirements for server environment support from our survey and workshops with data stewards and early career researchers, scalability, and extensibility (R5, R7, R8), we designed MaskAnyone as a web-based application accessible via standard web browsers. The architecture employs a Manager--Worker pattern to distribute tasks efficiently. The backend serves as the central hub for user interactions and job management, while workers handle the computational heavy-lifting of video masking. This setup allows for scalable and extensible operations without limiting the application's ability to run locally. Figure -architecture illustrates this high-level architecture. The front-end uses React and TypeScript to create a single-page application (SPA) that interacts with the backend via an HTTP API. The backend is accessible through a Nginx reverse proxy. %and is built with Python for consistency with the worker component. Data persistence is managed through PostgreSQL RDBMS, although media files are stored directly within the file system. Worker processes, also implemented in Python, register with the backend and receive masking jobs via an HTTP API. Docker and docker-compose are used for orchestration, making the application easy to set up with minimal prerequisites.

## *Multimodal Masking Process*

The core functionality of MaskAnyone lies in its masking process, as depicted in Figure 2 below. The user initiates this by uploading a target video and selecting it within the interface, prompting the Masking Dialog (Figure -masking-presets). Users are offered three routes: select a predefined preset, choose a custom preset, or manually configure the masking options. If a preset is selected, users can either commence masking directly, satisfying the usability requirement (R2), or refine the preset via the Masking Configuration Dialog.

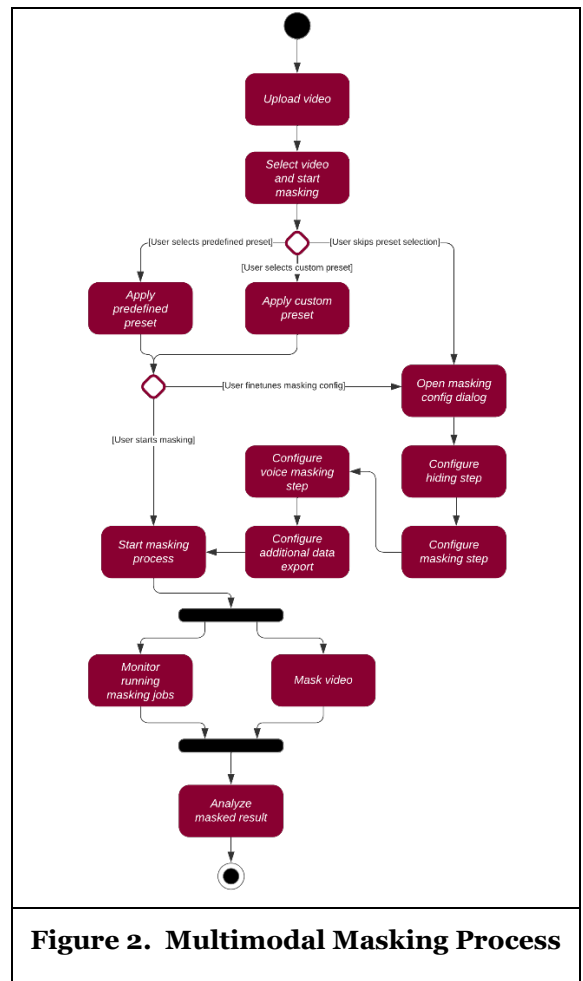

**Figure 2. Multimodal Masking Process**

The configuration process comprises four steps, offering users granular control over the masking. The first step provides options to control the hiding parameters for detected people and, optionally, the background. The second step involves specifying additional masking techniques to preserve important visual information. In the third step, decisions on audio masking options can be made, such as keeping, removing, or voice-converting the original audio. Finally, the fourth step allows for exporting additional data, such as kinematic information, either for more advanced analytics or external processing, in combination with audio-visual annotation tools (behavioral science) and NLP pipelines specific to the words in such masked videos. This workflow is designed to meet our articulated usability and flexibility requirements, offering a streamlined yet comprehensive masking solution.

### *User Interaction, Design Principles.*

The user interaction follows two design science principles: extensibility and scalability. The Manager--Worker pattern boosts scalability and extensibility. Users generate masking jobs in the backend; different job types can be added without altering the backend code. A new worker for the specific job type needs to be defined, adhering to established communication protocols. Conflicting dependencies for specific models are managed by creating unique Docker containers for each worker. The front-end integration of new masking methods is simplified using JSON Schema, allowing automated UI generation and job configuration validation. For scaling, MaskAnyone can add more worker processes and split large video files into multiple jobs. The primary bottleneck is the backend, which handles all user and worker interactions. While the backend can handle dozens of concurrent users and workers on robust hardware, further scaling may require partitioning it into multiple services. Significant extensions, like new categories of masking methods, may necessitate codebase modifications.

## *Video Masking Strategies*

As outlined in our GitHub repository, video masking in MaskAnyone involves two main aspects: hiding and masking. Hiding focuses on de-identification, while masking aims to preserve as much information in the video as possible. The toolkit offers a range of predefined options (i.e., strategies) for both categories.

**Video Masking Strategy: Hiding (S1)**

MaskAnyone employs two primary methods for detecting people in videos: YOLOv8 and MediaPipe. YOLO offers multiple pre-trained models with varying levels of accuracy and computational demand, including a specialized face-detection model. MediaPipe provides pose detection and an internal person mask. Users can adjust parameters like confidence thresholds for both methods to fine-tune utility and privacy tradeoffs[5]. After successful person detection, MaskAnyone offers several hiding techniques, visualized in Table 2 below:

| Strategy | Description | Effectiveness | Illustrative Results |
|---|---|---|---|
| Blackout | Completely blacks out the detected person, removing most identifiable information if the mask is accurate. | High, provided the mask is accurately applied. | 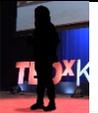 |
| Blur | Applies Gaussian blurring at a user-controllable blur intensity level. | Moderate, may be limited depending on blur intensity. | 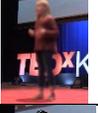 |
| Contours | Uses Canny edge detection to preserve essential contours while eliminating finer details. User-controlled detail. | Moderate, balances privacy with utility. | 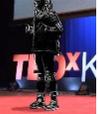 |
| Video Inpainting | Attempts to remove the person entirely, using surrounding data to fill in the background. | High, avoids residual artifacts but is computationally intensive. | 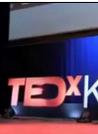 |
| **Table 2. MaskAnyone Strategy Elements (Hiding Approach)** | | | |

Note that video inpainting methods are resource-intensive and may require powerful hardware for timely execution. The integrated STTN model is limited to fixed resolutions, but alternative models supporting different resolutions are available.

**Video Masking Strategy: Masking (S2)**

MaskAnyone provides various masking options to preserve key information while maintaining privacy. These are summarized in Table 3 below and include:

| Strategy | Description | Effectiveness | Sample Illustration |
|---|---|---|---|

---

[5] For UI screenshots and videos of the different masking techniques, see the .mp4 files at https://anonymous.4open.science/r/Privacy4MultimodalAnalysis-57D4/README.md

| | | | |
|---|---|---|---|
| Skeleton | Captures body pose using 33 key points, rendering a skeleton overlay to preserve motion context. | High for motion context preservation. | 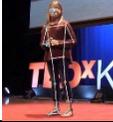 |
| Face-Mesh | Extracts and renders 478 facial landmarks, preserving facial features and expressions when faces are adequately sized in the frame. | High for facial detail preservation. | 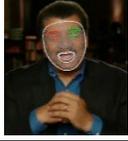 |
| Holistic | Combines pose, face, and hand landmark detection in a heuristic-based implementation. | Comprehensive for full-body interaction contexts. | A standalone program yet to be integrated into MaskAnyone. |
| Face Swap | Swaps the subject's face with a selected target face, maintaining expressions while hiding identity. | High for identity concealment while preserving expressions. | 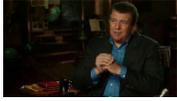 |
| Rendered Avatar | Utilizes Blender and the BlendARMocap plugin to transform motion coordinates into a 3D avatar. | High for anonymity, maintains motion fidelity. | 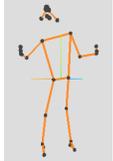 |
| Blendshapes Facial Avatar | Uses MediaPipe Face-Mesh to apply blendshapes on a 3D face, effectively preserving facial features. | High for facial expressions and detail. | 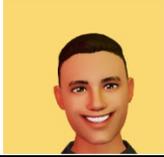 |
| **Table 3. MaskAnyone Strategy Elements (Masking Approach)** | | | |

**Voice Masking (S3)**

Voice masking is a crucial extension of MaskAnyone's video masking capabilities. We offer three primary approaches to audio handling, as enumerated in Table 4 below:

| Strategy | Description | Effectiveness | Sample Demonstration |
|---|---|---|---|
| Preserve | Retains the original audio and is suitable only for video masking. | Limited privacy, maintains original audio. | N/A |
| Remove | Eliminates all audio for maximum privacy. | High for privacy, low for utility. | N/A |
| Switch | Transforms the original voice to a target voice. | Balances privacy with utility and engagement. | N/A |
| **Table 4. MaskAnyone Strategy Elements (Audio)** | | | |

These features were implemented using MoviePy for voice extraction, RVC for voice conversion, and FFmpeg for audio-video merging. Various pre-trained target voices are available, and the system is extensible. We are also exploring other voice masking techniques, such as those from the Voice Privacy Challenge. End-users are advised to manually evaluate the privacy level/requirements before sharing.

# Preliminary Evaluation - Towards a Robust Evaluation Framework

Evaluating video masking tools is crucial for determining the ethical benchmarks reached by behavioral science research. However, masking aims to balance privacy preservation and utility retention. Therefore, robust masking must be evaluated on both accounts, and we propose an automatic evaluation methodology. Automated evaluations offer the advantages of standardization and scalability, which is particularly important for defending against computational re-identification methods. We also consider human evaluations an important complementary form of assessment that caters to behavioral researchers' specific needs and accounts for the subjective nature of privacy and utility. Our techniques may still be deemed effective in scenarios where the trade-off between privacy and utility is acceptable to human evaluators, even if not to machines.

## *Automatic Evaluation*

### Video Evaluation

Utility: Our evaluation strategy is divided into privacy preservation and utility retention. For utility, we focus on a specific use-case: emotion classification. We apply our face-swapping technique on selected videos with precise headshots and then use an existing emotion classification model to categorize emotions in these videos. We introduce an agreement score that measures the concordance between the emotional states classified in the original and anonymized videos. This score serves as a proxy for utility retention, with a high agreement indicating practical preservation of emotional cues. We sourced five images and three videos for this preliminary evaluation, with subjects explicitly facing the camera. Face-swapping was also applied to generate 15 test samples. The insights from our survey and user studies underscore the importance of the usability of the toolkit. Moreover, the issues raised by the participants relate to the utility of the data after it has been masked, depending on the research context and strategies selected to preserve utility. At the same time, we learned that fully anonymizing videos is not 100% guaranteed; however, the tradeoff can be balanced sensibly as more researchers can reproduce existing research when they can access the data. This suggests avenues for further refinement of our proposed methodology.

Privacy: We categorize our techniques into hiding and masking methods for privacy preservation. Hiding techniques primarily focus on object detection models, such as YOLOv8, to identify silhouettes, bounding boxes, and facial features. Their efficacy is usually assessed using the mean average precision (MAP) metric. In our case, the face-specific model achieved a MAP of 38%, and person detection models yielded MAP scores ranging from 37.3% to 53.9%, depending on the version used. Masking techniques, on the other hand, are evaluated through re-identification tasks. Due to the complexity of these tasks, a specialized dataset and model are often required. We fine-tuned a Vision Transformer and triplet-loss-based Re-ID model using a subset of the Celeb-A dataset. Preliminary results indicate a low precision score of 0.0017%, suggesting practical privacy preservation. However, these results also emphasize the need for further research to validate whether the low precision score indicates strong privacy or a model limitation.

### Audio Evaluation

We have not performed a voice masking evaluation, but we outline a planned methodology for future research focusing on utility and privacy. Utility assessment in voice masking can be complex, as it varies by use case. However, general metrics like Word-Error Rate (WER) and Pitch Correlation could be considered. WER can be computed using an automatic speech recognition system (ASR) like ESPnet or OpenAI's Whisper. Pitch Correlation, significant for fields like psycholinguistics, can be measured using the Pearson Correlation Coefficient between the input and output signals. The Librispeech dataset, widely used in ASR evaluations, can be harnessed for this purpose. For privacy, we propose a re-identification or automatic speaker verification (ASV) approach for privacy evaluation, using Equal Error Rate (EER) as the metric. EER effectively gauges the ability to re-identify individuals in audio data and is implemented in popular toolkits such as Kaldi. While the Librispeech dataset could potentially be used for privacy evaluation, the suitability of other datasets should also be explored.

*Human Evaluation*

Human evaluation is pivotal in privacy and utility preservation and is the most challenging part of such design science projects. We conducted a user study involving Ph.D. students(n=15). The students were divided into two groups to serve as both control and treatment groups. Initially, each group was shown unmasked celebrity videos to establish a recognition baseline. They were then shown masked videos and asked to identify the individuals. One unmasked video was included to assess the impact of initial biasing. The results are also visualized here[6] on the GitHub page.  Notably, face-swapping techniques significantly reduced re-identification scores. For instance, none could correctly identify Jackie Chan, possibly due to the choice of the replacement face or cultural factors. Priming also had a notable impact, as demonstrated by the difference in identification rates for the same video of Steph Curry when presented as either masked or unmasked. These findings underscore the necessity for further research to validate and improve upon our masking techniques. We also conducted two more user feedback studies with researchers(n=16) at another European university. We can report that data stewards and early career researchers registered interest in subsequently participating in a more robust evaluation of the toolkit. We have also taken some of their feedback to streamline the masking strategies and added a login/database management option to manage access rights between researchers in the same faculty/lab to masked data.

# Discussion

Maintaining privacy and utility in video masking is challenging and context-dependent. Different masking and hiding techniques may excel in privacy but limit utility in specific research settings. For this reason, we have developed MaskAnyone, a platform that combines masking and hiding techniques to tailor to the specific needs of researchers and other users. This masking tool also invites a discussion and a need to validate what combination of features constitutes a certain level of privacy and utility. In this paper, we have also proposed novel methods for evaluating those dimensions. We hope to contribute to such evidence-based measures and become better situated within ethical regulatory frameworks like GDPR. A vital issue is that it is unclear what counts as a proper reduction of identifiable information or anonymization.  Human evaluation experiments might be insightful in this regard, but they come with limitations, too. Using celebrities to assess identifiability leverages background knowledge but likely inflates re-identification scores relative to real-world use cases. This is because researchers who share data often do not know the persons who are part of the original data recorded; the risk of identification is low. Legislation in the European Union is also evolving on this matter, such that although identification might be possible in principle if, in practice, data users are not likely to ascertain someone's identity, then a reduced risk of privacy violation is implied (EUR-Lex 2023). Further discussion points for our proof-of-concept evaluation methods include that providing a larger pool of identification choices may further reduce the likelihood of correct identification, as evidenced by our last test video, where no options were given. Important considerations include the cost and the typically small sample sizes of the resulting Human evaluations. Finally, specific to our experiment, our small sample size limits the generalizability of our findings.

*Future opportunities and ethical challenges*

However, when considering these preliminaries, human evaluations and automated techniques of the sort we have proposed may be crucial to ensure a multi-faceted assessment of masking techniques. MaskAnyone offers a range of options to researchers, allowing customization based on specific needs for privacy and utility. However, several untapped scenarios offer avenues for future development. The toolkit can expand on voice customization features for multiple individuals in a single video and allow for masking self-defined objects or persons. Moreover, since the toolkit can effectively anonymize video call recordings offline (see, e.g., link), it could also be implemented as part of the recording process. Such an online audio-visual masking would ensure that identifiable data is never stored outside the masking process). This potential expansion of the utility of MaskAnyone could be significant in CCTV contexts where a lot of audio-visual data is collected that is not necessarily for purposes of identification and widespread masking while preserving other information (number of people, types of activities, group forming) could change surveillance practices for the better. These uncovered use cases and possible extensions indicate room for the tool's evolution to meet diverse privacy and utility needs. And the best part? Since MaskAnyone is open source, it invites you, the audience, to shape its potential further. This first version of MaskAnyone has several limitations. Factors such as lighting, camera angles, and the presence of multiple subjects can influence anonymization accuracy and sometimes lead to identity leaks. Additionally, fine details like micro-expressions may also be compromised. Another aspect to consider is the toolkit's resource intensiveness, which could limit its accessibility. High computational demands and a 50GB storage requirement could pose barriers, although a lightweight installation is available to mitigate these issues. Scalability could be an issue, especially when handling large datasets or multiple users, leading to performance bottlenecks. Ethical and legal considerations add further complexities. Potential misuse of technology and compliance with data protection laws are issues that we are looking to address in close engagement with data stewards and ethics personnel. Several technical enhancements are planned to refine masking techniques and explore more robust machine-learning solutions. Such enhancements will focus on making the toolkit more robust, interactive, and user-friendly, and we are exploring incorporating identity and access management tools like Keycloak. We have ongoing collaborations with researchers and data stewards so that the toolkit's development follows real-world needs and to explore how masking toolkits can become integrated with ethical data archiving and sharing procedures. We also plan to enhance the data utility aspect of the toolkit by incorporating modules for multimodal behavior analyses and classification (multimodal analytics), which may include gesture detection (Ripperda et al. 2020) and analyses(Dang and Buschek 2021; Zeng et al. 2023), and a summary of prosodic markers in the audio before voice masking (ref). One other line of application of our toolkit that is timely relates to analyzing deep-fake videos and AI-generated content, seeing the developments of SORA (Ho et al. 2022; openai 2024; Yan et al. 2021)

---

[6] https://anonymous.4open.science/r/Privacy4MultimodalAnalysis-57D4/README.md

## Conclusion

Audio-visual data containing human subjects are central to behavioral sciences and linguistics research, offering rich insights into human behavior and multimodal language use. (Dale 2008; Gregori et al. 2023; Kim and Adler 2015; Linzen 2020) However, while central to these fields, audio-visual data also raises ethical and privacy risks. In this context, GDPR and other legislatures play a pivotal role, providing regulatory frameworks that researchers must navigate to balance privacy concerns with the reproducibility, re-usability, and broader research utility of the data collected. Drawing on Design science, we envision how problem-centered artifacts, as instantiated in toolkits like MaskAnyone, may become integrated with ethical application procedures for audiovisual data archiving and research at universities and research institutes. This could mean that less fully identifiable audiovisual data is stored than is strictly necessary for archiving purposes. It could also mean that more audiovisual data can be safely shared with other researchers as masking minimizes privacy risks. Furthermore, we envision that research groups also integrate masking toolkits into their communication with other researchers at conferences, for example. Though there is no systematic study on this, it appears not uncommon for researchers to either not exemplify the audiovisual data that their research is based on due to privacy issues or a set of different kinds of video editing tools (e.g., g, Adobe Premiere Pro, Powerpoint) are used to add a static box as a mask of the person's face which is a suboptimal solution given the technological advancements in computer vision. In sum, we believe that masking toolkits like MaskAnyone can significantly improve the ecology surrounding research archiving and open science by providing a standardized set of masking and hiding strategies. Proper integration into research practices would mean that ethical review boards at research institutes could help advise or review what masking strategy is most optimal for a particular research context. Developing and adopting sophisticated data masking tools like MaskAnyone is vital for advancing ethical research practices in the digital age. These tools support compliance with stringent privacy regulations and foster a culture of responsible data sharing, thereby enhancing the integrity and utility of research in behavioral sciences and linguistics. The potential for such tools to be reviewed and recommended by ethical review boards further underscores their importance in aligning technological capabilities with ethical research standards. Our toolkit serves a diverse range of users including journalists, activists, and lawyers, addressing both the potential for misuse, like creating deceptive deepfakes, and the need for privacy. We aim to balance these concerns by engaging with data stewards and researchers to refine the use of our masking tools, MaskAnyone, in ethical guidelines for socio-behavioral research. Standardizing such techniques in academic data management can enhance privacy protections, allowing for safer data sharing and alleviating concerns associated with presenting sensitive audio-visual data. This approach encourages the use of more effective masking methods over traditional, less secure techniques.

## Impact Statement

This paper presents MaskAnyone, a de-identification toolkit poised to significantly impact the fields of behavioral sciences, linguistics, and any social science research involving sensitive audio-visual data. By introducing advanced, user-friendly masking technologies, MaskAnyone is designed to address pressing ethical and privacy concerns that often impede the sharing and utilizing of audio-visual data in research. We anticipate this toolkit will enable broader, safer data-sharing practices in social science research. Early career researchers, Data stewards, and institutional review boards are the primary audiences who will benefit from this toolkit. They will find the artifact invaluable for augmenting ethical research, enhancing reproducibility, and fostering open science while balancing de-identification concerns with data utility. Additionally, the toolkit's design principles and underlying requirements serve as a model for developing similar tools across various data-sensitive fields, e.g., healthcare and surveillance, towards promoting a culture of ethical data use. Ultimately, this paper could guide policy-making and moral standards in research institutions, advocating for a balanced approach to privacy and utility in audio-visual research data management.

MaskAnyone Toolkit